# A PROBABILISTIC DEMAND SIDE MANAGEMENT APPROACH BY CONSUMPTION ADMISSION CONTROL

*Lorant Kovacs, Rajmund Drenyovszki, Andras Olah, Janos Levendovszky, Kalman Tornai, Istvan Pinter*

New generation electricity network called Smart Grid is a recently conceived vision for a cleaner, more efficient and cheaper electricity system. One of the major challenges of electricity network is that generation and consumption should be balanced at every moment. This paper introduces a new concept for controlling the demand side by the means of automatically enabling/disabling electric appliances to make sure that the demand is in match with the available supplies, based on the statistical characterization of the need. In our new approach instead of using hard limits we estimate the tail probability of the demand distribution and control system by using the principles and the results of statistical resource management.

## 1 Introduction

The main issue in electricity networks is keeping an almost perfect balance between electricity generation and consumption all the time. Balance between demand and supply is crucial since oversupply means waste of energy, while undersupply causes performance degradation of the grid parameters (e.g. phase, voltage level, etc.). Unfortunately the control of the supply side is almost impossible because of the large time constants of the fossil and nuclear plants; the only possibility is applying cost ineffective auxiliary generators. Additionally, in smart grids the percentage of renewable resources should be increased which gives rise to uncertainty in the generation side. Hence, the best way to keep the balance is to manage the demand side. Demand Side Management (DSM) means a new kind of challenge: system operators should control the power grid in local scale, which is possible by installing intelligent measurement devices (smart meters). However, as a new perspective, households can be controlled with the intelligent devices. The residential sector accounts for about 30% of total energy consumption [1] and contains time shiftable appliances in high number. The amount of consumption involved in direct control can eliminate the error between daily prediction based generation and actual demand. The spread of electric vehicles could mean an additional opportunity. In average cars are parked in Europe for more than 90% of the time [2]; hence, batteries of electric vehicles can serve as an extra storage capacity for the power grid.

In this paper we propose a new approach for short-time demand side management. The introduced method takes into account the probabilistic nature of the load by the aid of a consumption admission control. The algorithm enables/disables shiftable appliances and reshapes the probability density function (pdf) of the aggregate consumption.

### 1.1 Related work

Influencing the demand side in the context of Smart Grid electricity networks is usually referred to as Demand Side Management (DSM) or Demand Response (DR). Demand Response is a mechanism managing customer consumption in response to supply side conditions while Demand Side Management covers all the activities or programs undertaken by service providers to influence the amount or timing of electricity use. There are many solutions proposed to DR and DSM like direct control of smart appliances, pricing and load scheduling. Good references can be found about different DSM approaches in [3], [4] and [5]. With direct control system operators can remove the extreme values in electricity consumption (peak shaving) and encourage additional energy use during periods of lowest system demand (valley filling). The load control as a demand response strategy is presented in [6], where simulating (summer period, air-conditioning units) is conducted with two control algorithms. It takes into account users' comfort (via heuristic consumer utility metric) and uses binary on-off policies. Fairness is maintained by two scheduling algorithms: priority based and round robin. Results show that significant energy and cost savings can be achieved with the proposed algorithms.

To minimize the operating cost of a residential microgrid, a MILP model is proposed in [7]. Decision variables are used to model demand and also supply of both electrical and thermal energy. It covers solar energy, distributed generators, energy storages, and loads. A model predictive control scheme is proposed to iteratively produce a control sequence for the microgrid. The case study reveals the performance of minimum cost control by comparison with benchmark control policies. Results show savings in annual operating cost.

A DSM model with three layers is introduced in [8]. This model consists of admission controller, load balancer, and demand response with load forecaster modules. Whenever a user turns on an appliance, a request is sent to the admission controller. If the capacity is available and we are not in peak hours, then it accepts the request and initiates the operation of the appliance. If the appliance operation exceeds the capacity, then the request is rejected and forwarded to the load balancer. The task of the load balancer is to solve an optimization problem and to assign a future timeslot for the appliance to start later. A game-theoretic approach for residential energy consumption scheduling is proposed in [9]. It introduces a pricing mechanism which is based on a convex and increasing cost function. The authors present a distributed algorithm for optimization problem. Most of the proposed techniques in the literature consider fixed load curves. However most of the papers do not deal with randomness on the load side, an exception is [10], where uncertainty is considered as well. The authors propose a MILP optimization model, which performs scheduling. The adopted DSM model forecasts the load curve of the user from the previous knowledge of their energy usage. Additionally, real time pricing and inclining block rates are combined in the model for effective pricing. The optimization is multi-stage, as the information of the appliances is revealed over time, the schedule of the appliances is updated accordingly. Simulation results show efficiency by reducing total peak-to-average ratio and energy expenses of users.

The rest of this paper is organized as follows. The problem formulation and system model is described in Section 3. The concept of our proposed Consumption Admission Control algorithm is introduced in Section 4. The results are presented with discussion in Section 5. Finally the paper is concluded in Section 6.

## 2 Problem formulation and system model

A large number of appliances can tolerate some delay (e.g. executing the program of a washing machine at a later time). Additionally in the near future the spread of electric vehicles will mean a huge amount of elastic demand in the power system. As a result, there are (time) shiftable and non-shiftable demands in the system. On the other hand most of the devices show stochastic consumption behaviour (neither the start of use nor time of operation is known a priori), hence, only a statistical approach can efficiently solve the control task. The foundation of our approach is that, the new Consumption Admission Control algorithm can modify the pdf of a consumption unit (e.g. a household, street, city etc.) by the temporary enabling/disabling of shiftable appliances. From the system operators' perspective pdf close to Dirac delta function (meaning constant load) is ideal. However, we cannot reach the optimum, as a more realistic goal, we can keep the probability density function as narrow as possible, i.e. the mass of the pdf lies between a lower and an upper limit. For the sake of an even more realistic model, we allow the tail probabilities to be non-zero but smaller than a predefined probability.

In this paper, we assume that the service provider calculates and communicates these parameters governing the behaviour of a customer. Using these parameters, the subscriber's Smart Meter (SM) can enable/disable the appliances at a local level, resulting in a fully distributed solution to the problem. The parameters coming from the service provider that govern the CAC algorithm are as follows: a capacity upper limit ($C_{max}$) in every time slot, which is allowed to be exceeded by a small probability $p$. (In this paper we will concentrate on the upper limit $C_{max}$ and oversonsumption probability, however we plan to extend our approach by a lower limit $C_{min}$ and an underconsumption probality $r$ in our future work). The tail probability will be referred to as Quality of Service (QoS) parameter, because it can satisfy the overload of the grid to keep under a certain limit, and hence, the stability of the grid parameters such as frequency, voltage level, etc. The task of the SM is the admission control (enabling/disabling) of the appliances by such a way, that the probability distribution function (pdf) of the aggregate consumption satisfies the prescription of the service provider. (The cooperative attitude of the subscriber can be motivated by rewards.) The underlying model is depicted in Figure 1.

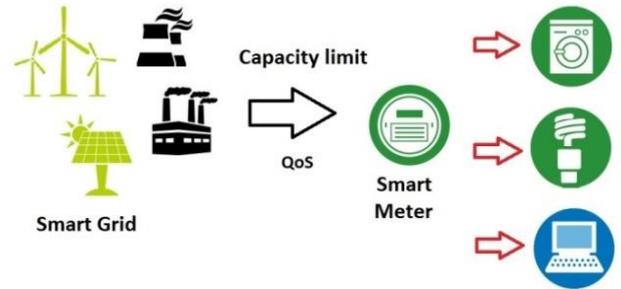

**Figure 1** The applied model

In the model all subscribers are assumed to have a Smart Meter. The SM has the following properties:
- the SM can communicate with the service provider and with the smart appliances;
- the SM can register the consumption statistics of the appliances (both smart and traditional ones);
- the SM can temporarily enable/disable appliances.

In the model stochastic and deterministic, shiftable and non-shiftable appliances are taken into consideration. (The devices executing a fixed program can be seen as deterministic). The defined categories and some examples of appliances are listed in Table 1.

**Table 1** Device categories used in the model

|  | *stochastic* | *deterministic* |
|---|---|---|
| *shiftable* | electric heating, air conditioner, refrigerator | washing machine, dishwasher |
| *non-shiftable* | lighting, vacuum cleaner | circulation pump |

In Figure 2 the measures used in the model are depicted. The admission control algorithm uses discrete time slots (denoted by k in Figure 3), in which the enabled/disabled status of the appliances and the system parameters (capacity limits and QoS) are supposed to be unchanged. (New consumption requests are supposed to be handled instantaneously). In all time slots there is a deterministic component of the consumption and as well as a stochastic one.

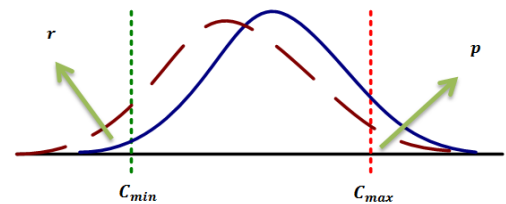

**Figure 2** Illustration of the original and modified pdf of the aggregate consumption and the free parameters that govern the algorithm

The stochastic part is described by its estimated (or calculated) probability distribution function. The maximum ($C_{max}$) and minimum ($C_{min}$) capacity limits can be changed in every time instant by the service provider. $C_{sys}$ builds a natural upper limit (i.e. lines and fuses) on $C_{max}$.

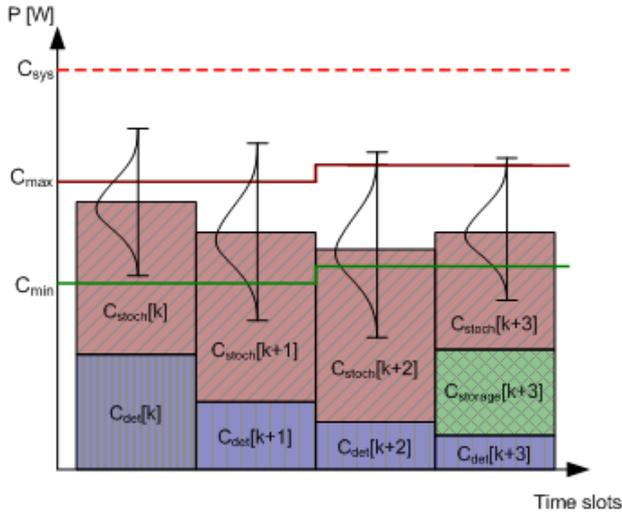

**Figure 3** Measures used in the model

## 3 Consumption Admission Control Algorithm

The decision to enable or disable an appliance in the system is carried out by the Consumption Admission Control (CAC) algorithm. As mentioned in Section 2 the aim of the algorithm is to sharpen the shape of the pdf of the aggregate consumption of a customer resulting near constant load in the time domain. The Smart Meter calculates the aggregate pdf from the individual pdf-s of the appliances. The individual pdf can be communicated to the SM by smart appliances, or it can be measured in the case of traditional ones. This concept was originally applied for Call Admission Control for ATM communication networks [11].

In this paper the following mathematical model will be used: Let $X_j$ denote the random variable of the consumption of the jth appliance, while

$$X = \sum_{j=1}^{N} X_j \quad (1)$$

is the aggregate consumption random variable and $N$ is the number of enabled appliances.

In the case of a new incoming consumption demand, the CAC checks whether the inequality (2) holds for the enabled plus the incoming appliance

$$Pr(X \geq C_{max}) \leq p \quad (2)$$

where $Pr$ denotes probability of an event, and $p$ is the probability limit of overconsumption. Therefore, CAC keeps the upper tail probability of the aggregate consumption under the limit $p$.

The probability of overconsumption $Pr(X \geq C_{max})$ can be calculated based on the probability density function $f_X(x)$ of the aggregate consumption. The pdf of the aggregate consumption can be calculated analytically by the convolution of the individual pdfs of all appliances:

$$f_X(x) = Pr\left(\sum_{i=1}^{M}\sum_{j=1}^{n_i} X_{ij} = x\right) =$$
$$= f_{X_{11}}(x) * f_{X_{12}}(x) * f_{X_{13}}(x) * \ldots * f_{X_{Mn_i}}(x) \quad (3)$$

where $M$ is the number of appliance classes, and $n_i$ is the number of appliances in class $j$, and $\sum_{i=1}^{M} n_i$ is the total number of enabled appliances (An appliance class means a set of appliances that have the same statistical descriptors). Considering deterministic ($X^{det}$) and stochastic ($X^{stoch}$) appliances in the model, we can write the inequality:

$$Pr(X^{stoch} + X^{det} \geq C_{max}) \leq p, \quad (4)$$

$X^{det}$ is a constant value so the probability can be expressed as

$$Pr(X^{stoch} \geq C_{max} - X^{det}) \leq p; \quad (5)$$

The lower limit can be checked by the same manner as the upper limit. If the probability of underload is higher than $r$, the goal can be expressed as

$$Pr(X < C_{min}) \leq r; \quad (6)$$

### 3.1 Estimation of the probability of overconsumption

The convolution operation in (3) can be very time consuming in the case of high number of appliances and/or classes, so it is suggested to estimate the probability in terms of inequalities [12] of Large Deviation Theory (LDT) bounds, such as Markov, Chebisev, Bennett, Hoeffding and Chernoff upper bounds. The estimation of overconsumption can be derived from the calculation of the following upper bound:

$$Pr(X \geq C_{max}) \leq \widehat{U}(X, C_{max}) \leq p \quad (7)$$

where $\widehat{U}(X, C_{max})$ is the bounding method on the tail probability. Because of the independence of the $X_{ij}$ random variables, the expected value can be expressed as

$$\mu = E\{X\} = \sum_{i=1}^{M}\sum_{j=1}^{n_i} \mu_{ij} \quad (8)$$

and variance as

$$\sigma^2 = E\{(X-\mu)^2\} = \sum_{i=1}^{M}\sum_{j=1}^{n_i} \sigma_{ij}^2 \quad (9)$$

The most widely known, Markov's inequality needs only expected value to give an upper bound for the probability that the non-negative $X$ random variable is greater than or equal to some positive constant ($C_{max}$ in our case) :

$$Pr(X \geq C_{max}) \leq \frac{\mu}{C_{max}} \quad (10)$$

Inevitably the advantage of Markov's inequality is its simplicity, but it is not a tight upper bound.
Chebysev's inequality

$$Pr(X \geq C_{max}) \leq \frac{\sigma^2}{(C_{max} - \mu)^2} \qquad (11)$$

is also simple, but it is also not a tight upper bound.

Hoeffding's inequality is an exponentially decreasing upper bound, which results in a tighter estimation even far from the expected value compared to Markov's and Chebysev's inequalities. It is also based on the expectation that $X_{ij}$ random variables are independent and additionally $X_{ij}$ variables have upper and lower bounds: $x_{ijmin} \leq X_{ij} \leq x_{ijmax}$. Hoeffding's inequality [13] can be expressed as:

$$Pr(X \geq C_{max}) \leq exp\left(\frac{-2(C_{max} - \mu)^2}{\sum_i \sum_j (x_{ijmax} - x_{ijmin})^2}\right) \qquad (12)$$

From (12) it is clear that with the increase of $C_{max}$, the upper bound decreases in an exponential rate.

Bennett's inequality gives exponentially decreasing upper bound like Hoeffding's, which assumes bounded input random variables $|X_{ij}| \leq x_{max}$, and it is formulated in the following form [14]:

$$Pr(X \geq C_{max}) \leq$$
$$exp\left(-\frac{\sigma^2}{x_{max}^2} \cdot h\left(\frac{(C_{max} - \mu) \cdot x_{max}}{\sigma^2}\right)\right) \qquad (13)$$

where $h(u) = (1 + u) \log(1 + u) - u$. Bennett's inequality needs additional statistical information compared to Hoeffding's, the standard deviation of appliances ($\sigma_{ij}$) and maximum value ($x_{ijmax}$).

Chernoff's inequality is also an exponentially decreasing upper bound [15]:

$$Pr(X \geq C_{max}) \leq exp\left(\sum_{j=1}^{N} \mu_j(s^*) - s^* C_{max}\right) \qquad (14)$$

where $\mu_j(s) = \lg E\{e^{sX_j}\}$ are the so called logarithmic momentum generating functions and $s^*$ is the parameter that satisfies the possibly tightest bound:

$$s^* : \inf_{s>0} \sum_{j=1}^{J} \mu_j(s) - s C_{max} \qquad (15)$$

When a new demand of a shiftable appliance appears, enabling or disabling will be calculated using one of the upper-bounds:

$$sgn\{p - \hat{U}(X, C_{max})\} = \begin{cases} -1, 0 \; Accept \\ +1 \; Reject \end{cases} \qquad (16)$$

here X is the aggregate consumption random variable containing the consumption of all enabled (both shiftable and non-shiftable) appliances plus the incoming one.

Another approach for estimating an aggregate pdf is based on the Central Limit Theorem (CLT)

$$Pr(X > C_{max}) \leq 1 - F_X(C_{max}) \qquad (17a)$$

$$F_X(C_{max}) \to \Phi\left(\frac{C_{max} - \mu}{\sqrt{\sigma^2}}\right) \qquad (17b)$$

where $F(x)$ denotes the cdf of $x$ and $\Phi(.)$ is the standard normal cdf. We must emphasize that CLT is not an upper bound on the tail probability. The speed of convergence of $F(x) \to \Phi(x)$ is the main question regarding the estimations based on the Central Limit Theorem. The absolute error of the CLT estimation $|F(x) - \Phi(x)|$ is decreasing towards the tails, but the relative error $|F(x) - \Phi(x)|/\Phi(x)$ is increasing [16].

## 4 Results and discussion

In order to have a clear picture about the performance of the Consumption Admission Control algorithm a simulation environment was established in MATLAB. We investigated the following aspects of the CAC algorithm:
- Relation of QoS ($p$) and empirical probability of overconsumption ($\tilde{p}$) in the case of different LDT bounds;
- Model complexity of load time series;
- Load shape modification made by CAC;
- Number of enabled appliances in the case of different LDT bounds and CLT;

Throughout our simulations we used stationary load time series to explore the statistical behaviour of the CAC algorithm. It is clear that the real benefit of the new algorithm comes to the fore in a nonstationary environment such a day or longer consumption period.

### 4.1 Relation of QoS and empirical probability of overconsumption

In this section we present our investigation regarding the relation of predefined QoS and empirical probability of overconsumption. The ratio of predefined QoS and empirical probability of overconsumption will be denoted by

$$k = \frac{\tilde{p}}{p} \qquad (18)$$

Using an upper bound on the tail probability leads to underestimation of the number of appliances to be enabled which results in $\tilde{p} < p$, i.e. $k < 1$; and vice-versa a lower bound results in $k > 1$. From the point of view of the service provider, $k < 1$ means guaranteed QoS, but causes spare capacities.

The following assumptions were made in the simulations:
- Load of appliances were modelled by two-state Bernoulli iid series of 50000 time instants;

- There is only one appliance class. (All appliances have the same statistical descriptors.)
- Number of appliances in the class is 400;
- The consumption demand of the temporarily disabled appliances are deleted.

The aim of the investigation was to measure the performance of different tail probability estimation methods plugged into the CAC in the case of different probability of ON state of the appliances ($p_{ON}$). Figure 4 and 6 depict the results in the case of $p_{ON}=0,1$ and $p_{ON}=0,5$, respectively.

The results in Figure 4 and 5 show that the empirical probability can almost meet QoS ($k = 1$) when the tail probability is exactly calculated from the analytical aggregate pdf (see (3)). There is only a small deviation, $k = 0,4 \ldots 0,6$ in the case of small probabilities ($10^{-5} \ldots 10^{-4}$) due to the difficulty of measuring rare events in the case of Monte Carlo simulations.

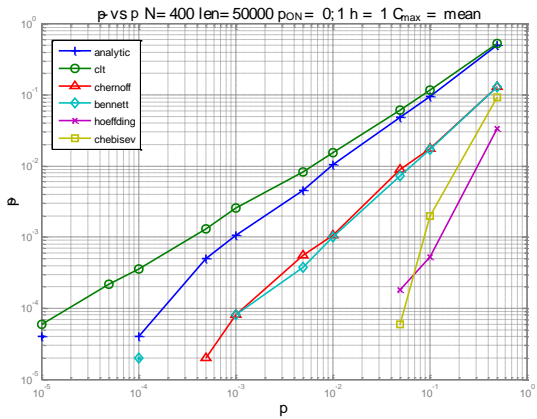

**Figure 4** $\tilde{p}$ vs $p$ for different bounds ($p_{ON}=0,1$)

Using Chernoff's and Bennett's inequalities the CAC algorithm sets with one order of magnitude lower the ratio $k$ regardless of $p_{ON}$, which results only in an acceptable decrease of the number of accepted appliances (for details see Section 4.4).

Applying Hoeffding bound leads to results which highly depend on the $p_{ON}$ value. Applying Chebisev and Markov bound lead to poor results regardless of the $p_{ON}$ values. The performance of CLT based CAC is close to the analytic calculation ($k = 1 \ldots 3$). Note that CLT is not an upper bound on the tail probability. As a consequence $k > 1$ values can occur.

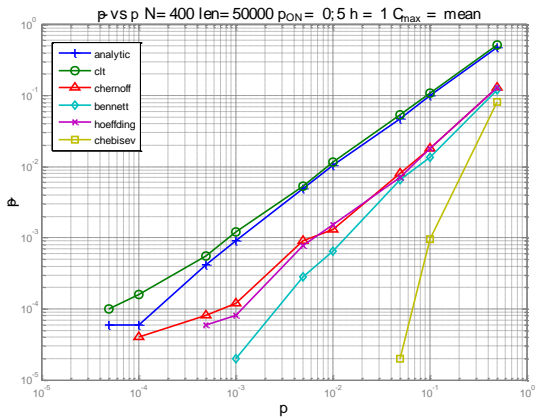

**Figure 5** $\tilde{p}$ vs $p$ for different bounds ($p_{ON}=0,5$)

## 4.2 Model complexity of load time series

The CAC algorithm needs appliance level statistical information, therefore, load time series in our simulations are generated with the bottom-up approach, i.e. the aggregate time series are built up from appliance level consumption time series. We used different appliance-level models in the simulations:
- Bernoulli iid;
- First Order Markovian;
- Higher Order Markovian.

In all the tree cases two-state (ON/OFF) models were used. Bernoulli iid is not a realistic consumption model, its aim is to prove the CAC concept. It requires only measuring the probability of the ON state and the maximum value of the consumption. A more realistic, widely used model is the First Order Markovian model [17]. This model can be described by a transition probability matrix. As the most realistic model among the three approaches we applied the distributions of the holding times for ON and OFF states separately which leads generally to a Higher Order Markovian (HOM) model. The benefit of HOM models is the capability to model long range dependence between samples, which is a usual property of real load time series. In Figure 6 examples of iid and HOM time series can be seen. In all the cases our models were fitted to measured data coming from the REDD DataSet [18]. The DataSet contains appliance level power data for 6 homes for several weeks with sampling time of 3 seconds.

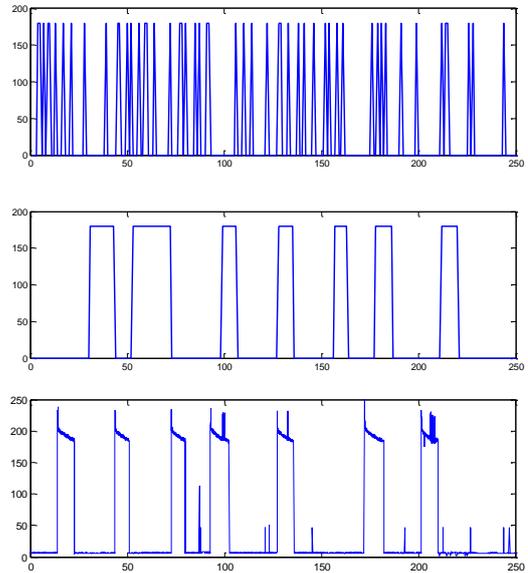

**Figure 6** iid Bernoulli (top) and HOM (middle) model and original measurement of a refrigerator (bottom)

The CAC algorithm descripted by equations (3), (10)-(14) assumes iid appliance load time series. It is an important question, how complex time series models affect CAC. Figure 7 demonstrates that there is only a slight performance degradation even with the HOM model (400 pieces of microwave ovens with ON probability of 0,0160; simulation length is 50000 time instances).

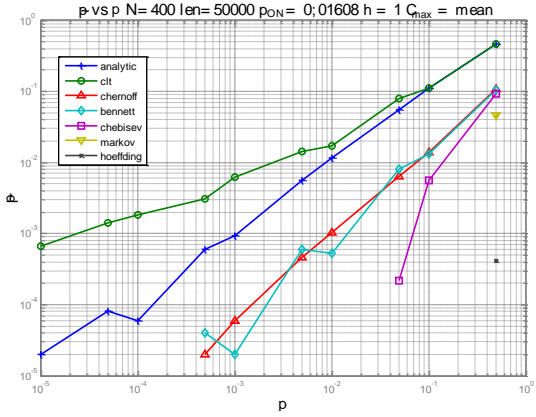

**Figure 7** $\tilde{p}$ vs $p$ with microwave oven HOM model

However, in the case of small probabilities ($10^{-5} \ldots 10^{-4}$) CLT results in a higher ratio ($k = 10 \ldots 20$), Chernoff and Bennett remains almost in the same range ($k = 0{,}1$) like in Figure 5 and 6.

**4.3 Load shape modification made by CAC**

However the basic mathematical idea of our CAC is to limit the over- and underconsumption probability, the direct objective of DSM methods is expressed as load shape modification in the time domain (for instance by valley filling and peak clipping). The CAC algorithm, as stated before, forms the pdf of the aggregate consumption towards the Dirac-delta function, which is equivalent to constant load in the time domain. In this section we demonstrate the effectiveness of the CAC algorithm regarding load shaping. Assumptions are:
- The consumption demand of the temporarily disabled appliances is deleted;
- Selection of the appliances to be temporarily disabled is based on random selection which guarantees fairness;
- One appliance class;
- All appliances are of shiftable stochastic type.

In Figure 8 the original aggregate consumption time series and the modified one can be seen. From the figure one can see, that this form of the algorithm does not yield almost any load shaping. Our hypothesis was that the treatment of consumption demand of the temporarily disabled appliances (which is referred to as scheduling strategy) plays key role in the algorithm to perform load curve modification. To prove this, we changed the scheduling strategy in the CAC to a so-called one-step strategy.

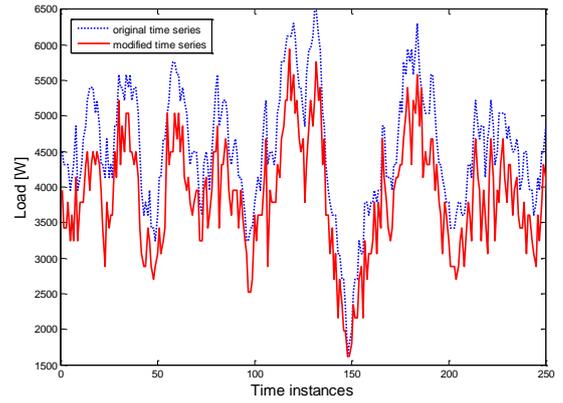

**Figure 8** Load shape modification ability of the CAC algorithm

The one-step scheduler (Figure 9) is an alternative method to handle the disabled appliances. In this case our assumptions are:
- The one-step scheduler shifts the consumption of temporarily disabled appliance with one time instant;
- It guarantees that the sum of the consumed energy remains the same after the modification of the load curve;
- Selection of the appliances to be temporarily disabled is based on random selection which guarantees fairness;
- One appliance class;
- All appliances are of schiftable stochastic type.

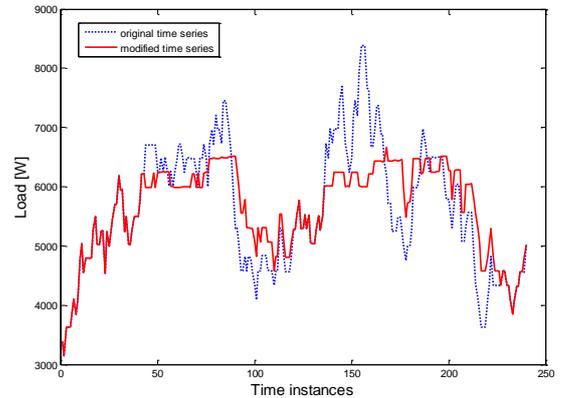

**Figure 9** Load shape modification with one-step scheduling

It is clear that the CAC with one-step scheduler is able to modify the load shape (red curve on Figure 9, which is closer to constant). The Load Factor (LF) is increased from 0,6718 to 0,8463 (LF is a widely used measure of the efficiency of electric energy usage, and calculated as $LF = \frac{average\ load}{maximum\ load}$). Based on the results, we are planning to investigate more sophisticated scheduling methods in our future work.

**4.4 Number of enabled appliances in the case of different LDT bounds and CLT**

In the CAC algorithm the scheduler disables shiftable appliances if the aggregate consumption exceeds the $C_{max}$ upper limit with a higher probability than it is allowed by $p$. The task in this step is to determine the number of appliances to be enabled in each appliance class so that the QoS must be satisfied. Figure 10 depicts the number of enabled appliances in the case of different LDT bounds and CLT, assuming:
- One appliance class modelled with Bernoulli iid model ($p_{ON} = 0,1$);
- 400 appliances.

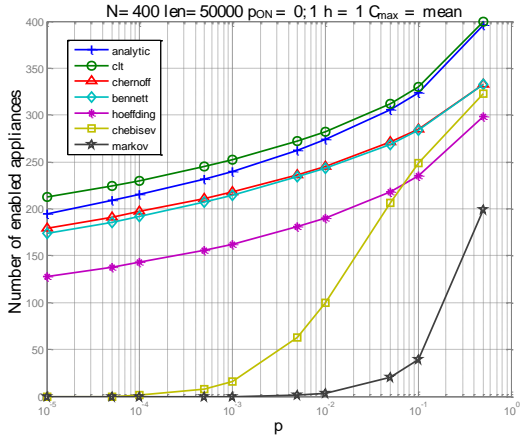

**Figure 10** Number of enabled appliances vs $p$

The number of enabled appliances is a monotonously increasing function of $p$ in the case of one appliance class (Figure 10). Estimation of the probability of overconsumption applying LDT bounds lead to lower number of enabled appliances compared to the analytically calculated value for all $p$ values. Applying LDT bounds, as stated before, causes spare capacities in the system. CLT is not a bound, so it can lead to values higher than 100%, which means breach of contract. The exact percentages of enabled appliances (with analytically calculated value as the reference) are collected in Table 2.

**Table 2** Percentage of enabled appliances

|  | $p < 10^{-3}$ | $p > 10^{-2}$ | QoS guaranteed |
|---|---|---|---|
| **Analytic** | *100% (reference)* | *100% (reference)* |  |
| **CLT** | 105% | 101% | no |
| **Chernoff** | 92% | 88% | yes |
| **Bennett** | 91% | 88% | yes |
| **Hoeffding** | 80% | 75% | yes |
| **Chebisev** | 0% | 50-80% | yes |
| **Markov** | 0% | 10-50% | yes |

In the case of two or more appliance classes, the CAC algorithm can decide to enable different combinations of appliances (Figure 11 and 12, where green colour indicates the allowable set of appliances, red colour indicates the combinations when the QoS is not satisfied). Assumptions are:
- Two appliance classes modelled with Bernoulli iid model (100 appliances in each classes);
- The tail probability is exactly calculated from the analytical aggregate pdf;

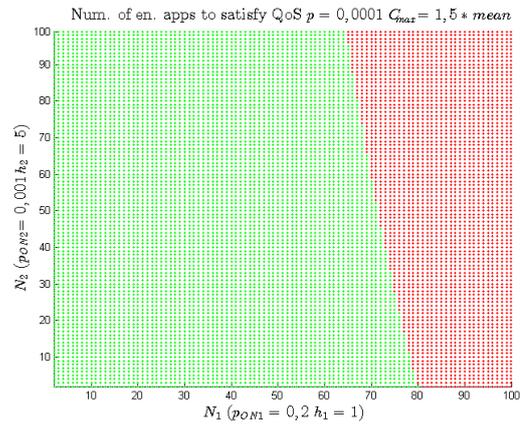

**Figure 11** Number of enabled appliances, one class ($h_2 = 5$)

Figure 11 shows that the decision curve is slightly nonlinear and convex, but with other parameters (Figure 12) it can be highly nonlinear and even non-convex. We can state that the two decision regions are generally not linearly separable.

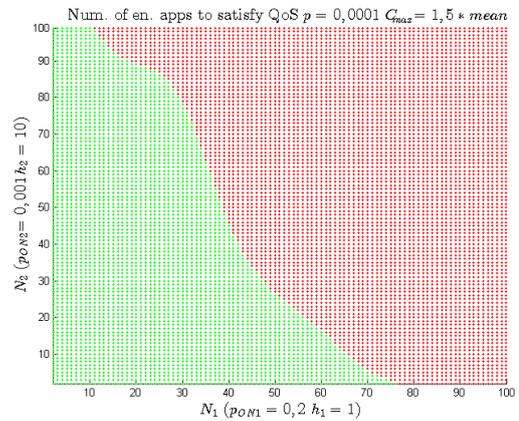

**Figure 12** Number of enabled appliances, one class ($h_2 = 10$)

The separator curve depends on the different LDT bounds and CLT applied in CAC. The next two figures (Figure 13, 14) depict the investigations regarding the number of enabled appliances in the case of different tail probability estimation methods. Assumptions are:
- Two appliance classes modelled with Bernoulli iid model (100 appliances in each classes);
- As a reference, the tail probability is exactly calculated from the analytical aggregate pdf.

In the first experiment (Figure 13) $p_{ON1}=0,2$ and $p_{ON2}=0,001$; and ON values $h_1=1W$ and $h_2=5W$. In the second experiment (Figure 14) the difference is only $h_2 = 10W$. The performance degradation is smaller in the first case when the difference between ON values ratio $h_2/h_1$ is not too large. In the case of higher $h_2/h_1$ ratio (Figure 14) the separator curves lie far to each other causing severe performance degradation.

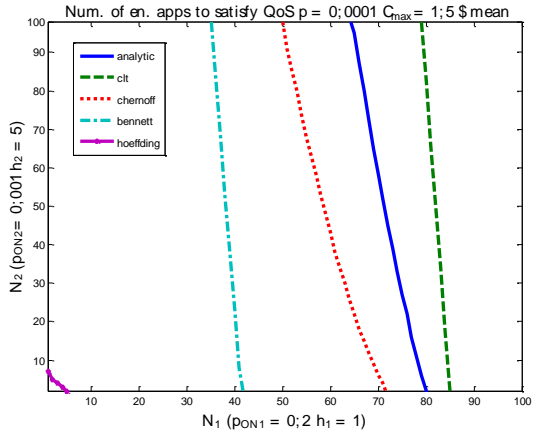

**Figure 13** Number of enabled appliances, two classes ($h_2 = 5$)

In the latter case (Figure 14) the exact separator is non-linear and non-convex but this fact is not reflected by the estimation methods.

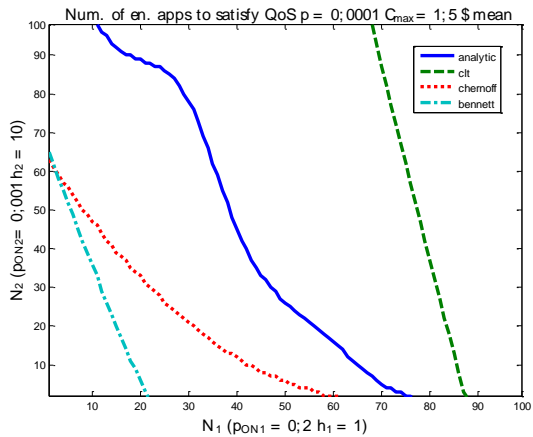

**Figure 14** Number of enabled appliances, two classes ($h_2 = 10$)

In Table 3 and 4 the number of enabled appliances can be seen for certain parameters.

**Table 3** Percentage of enabled appliances $h_2 = 5$

|  | **Analytic** | **CLT** | **Chernoff** | **Bennett** |
|---|---|---|---|---|
| **N$_2$=100** | 64* | 80 | 50 | 35 |
|  | *100% (reference)* | 110% | 91% | 82% |
| **N$_2$=50** | 71 | 82 | 58 | 38 |
|  | *100% (reference)* | 109% | 89% | 72% |
| **N$_2$=0** | 80 | 85 | 72 | 42 |
|  | *100% (reference)* | 106% | 90% | 53% |

*$N_1$

**Table 4** Percentage of enabled appliances $h_2 = 10$

|  | **Analytic** | **CLT** | **Chernoff** | **Bennett** |
|---|---|---|---|---|
| **N$_2$=100** | 11* | 69 | 0 ($N_2$=60) | 0 ($N_2$=60) |
|  | *reference* | 152% | 54% | 54% |
| **N$_2$=50** | 38 | 77 | 10 | 6 |
|  | *reference* | 144% | 68% | 64% |
| **N$_2$=0** | 76 | 88 | 60 | 21 |
|  | *reference* | 116% | 79% | 28% |

* $N_1$

The performance decrease caused by the different LDT bounds is the smallest in the case of Chernoff bound but it is highly sensitive to the $h_2/h_1$ ratio. In the case of $h_2/h_1 = 5$ the utilization loss caused by Chernoff bound is 9-11%. In the case of $h_2/h_1 = 10$ it is 21-46%. CLT has near the same performance but in the experiments the number of enabled appliances is higher than the reference which causes breach of contract regarding the QoS criterion $p$. At the same time the computational complexity of CLT is substantially lower than of Chernoff bound and analytical convolution. As a result we recommend using analytical computation when it is possible. In the case of lack of time and importance of satisfying QoS, Chernoff bound comes to the fore. CLT has the lowest computational need and has quite good performance but cannot guarantee QoS criterion.

## 5 Conclusions and future work

In this paper a new statistical approach was proposed for managing the balance between demand and available supplies in smart grids. The smart meter of the subscriber performs the task of enabling/disabling of shiftable appliances based on two parameters, obtained from the supplier: upper capacity limit and allowable probability of overconsumption (QoS). The smart meter influences the probability distribution function of the aggregate consumption in order to keep the tail probabilities under a given threshold $p$. The new approach takes the uncertainty of the consumption into account, and furthermore it can work in a fully distributed manner, since the calculations can be performed in the smart meter. We conducted several simulations to evaluate the performance of the CAC. As a result the introduced Consumption Admission Control method is a promising candidate for demand side management in smart grid environment.

## 6 Acknowledgement

Our research was supported by the European Union and the Hungarian Republic through the project TÁMOP-4.2.2.A-11/1/KONV-2012-0072 – Design and optimization of modernization and efficient operation of energy supply and utilization systems using renewable energy sources and ICTs.

**Authors' addresses**

*Lorant Kovacs, PhD*
Kecskemet College, Faculty of Mechanical Engineering and Automation (GAMF)
Izsaki ut 10, Kecskemet, H-6000, Hungary
kovacs.lorant@gamf.kefo.hu

*Rajmund Drenyovszki, MSc*
Kecskemet College, Faculty of Mechanical Engineering and Automation (GAMF)
Izsaki ut 10, Kecskemet, H-6000, Hungary
drenyovszki.rajmund@gamf.kefo.hu

*Andras Olah, PhD*
Pázmány Péter Catholic University, Faculty of Information Technology
Práter utca 50/a, H-1083 Budapest, Hungary
olah.andras@itk.ppke.hu

*Janos Levendovszky, PhD*
Budapest University of Technology and Economics, Deptartment of Telecommunications,
Egry József u. 18, H-1111 Budapest, Hungary
levendov@hit.bme.hu

*Kalman Tornai,PhD*
Pázmány Péter Catholic University, Faculty of Information Technology
Práter utca 50/a, H-1083 Budapest, Hungary
tornai.kalman@itk.ppke.hu

*Istvan Pinter, PhD*
Kecskemet College, Faculty of Mechanical Engineering and Automation (GAMF)
Izsaki ut 10, Kecskemet, H-6000, Hungary
pinter.istvan@gamf.kefo.hu